\documentclass[final,5p,times,twocolumn]{elsarticle}

\usepackage{color}
\usepackage{graphics}

\newcommand{\veps}{\varepsilon}

\journal{Physics Letters B}

\begin{document}

\begin{frontmatter}

\title{Accelerating Universe with spacetime torsion but without dark matter and dark energy}%
\author{A.~V.~Minkevich}%

\ead{minkav@bsu.by, awm@matman.uwm.edu.pl}

\address{Department of Theoretical Physics, Belarussian State
University, Minsk, Belarus} \address{Department of Physics and
Computer Methods, Warmia and Mazury University in Olsztyn, Poland}

\date{\today}

\begin{abstract}
It is shown that cosmological equations for homogeneous isotropic models deduced in
the framework of the Poincar\'e gauge theory of gravity by certain restrictions on indefinite
parameters of gravitational Lagrangian take at asymptotics the same form as cosmological equations
of general relativity theory for $\Lambda CDM$-model. Terms related to dark matter and dark energy
in cosmological equations of standard theory for $\Lambda CDM$-model are connected in
considered theory with the change of gravitational interaction provoked by spacetime torsion.
\end{abstract}


\end{frontmatter}

\section{Introduction}

As it is well known, the notions of dark matter (DM) and dark
energy (DE) were introduced in order to explain observational
cosmological and astrophysical data in the framework of the
general relativity theory (GR). According to obtained estimations
the contribution of these invisible components to the energy of
the Universe is approximately equal  to 96\%. The origin problem
of hypothetical kinds of gravitating matter -- DM and DE -- is one
of the most principal problems of modern cosmology and gravitation
theory. Many attempts have been proposed with the purpose of
solving this problem (see \cite{1,2,3,4,5,6,7} and references
herein). In the frame of GR the DE (or quintessence) as
gravitating matter with negative pressure provoking accelerating
cosmological expansion at present epoch is associated in many
works with vacuum energy leading to cosmological constant in
cosmological equations that is expressed in the title "$\Lambda
CDM$-model". In other works the quintessence is related to some
hypothetical fields. The DM is considered usually as connected
with some massive particles (WIMP), which appear in elementary
particles theory including various generalizations of Standard
Model and the search for which is realizing in many experimental
projects (see review \cite{7}). At the same time there is another
treatment that explains effects associated in the framework of the
standard theory with DE and DM. This treatment is connected with
the search for some generalization of Einstein gravitation theory,
where there is no DE and DM, and the corresponding effects are
connected with the change of gravitational interaction. At present
there are different approaches in this direction connected, in
particular, with extradimensional theories, $f(R)$ gravity, MOND
etc. Not always such theories are based on acceptable fundamental
physical principles.

The present Letter is devoted to the discussion of possible a
solution of DE- and DM-problems in the frame of the Poincar\'e
gauge theory of gravity (PGTG) (see \cite{8,9,10,11}), which is a
natural generalization of GR and which offers opportunities that
can solve the principal problems of Einstein gravitational theory
(see \cite{12, 13} and references herein). The PGTG is based on
well-known and acceptable physical principles including the local
gauge invariance principle, and it is the gravitation theory in
4-dimensional  physical spacetime with the structure of the
Riemann-Cartan continuum. Note that the PGTG is a necessary
generalization of GR, if one supposes that the Lorentz group,
which is fundamental group in physics, is included to gauge group
corresponding to gravitational interaction. At the first time the
simplest PGTG - the Einstein-Cartan theory \cite{14} - was applied
with the purpose to solve one of the problems of GR, the problem
of cosmological singularity in Refs \cite{15, 16}. However, the
possibilities of the Einstein-Cartan theory are limited. As a
natural generalization of Einstein-Cartan theory is the PGTG based
on gravitational Lagrangian ${\cal L}_{\rm g}$ including not only
scalar curvature but invariants quadratic in gravitational field
strengths - curvature $F_{\alpha\beta\mu\nu}$ and torsion
$S_{\alpha\mu\nu}$ tensors. By using sufficiently general
expression of ${\cal L}_{\rm g}$ regular isotropic cosmology
including inflationary cosmology was built and investigated in the
frame of PGTG (see \cite{17,18,19,20} and references herein). As
it was shown, the character of gravitational interaction by
certain physical conditions in the frame of PGTG is changed, and
in the case of usual gravitating matter with positive values of
energy density and pressure the gravitational interaction can be
repulsive \cite{13, 17, 19} that offers opportunities to solve
principal problems of GR, in particular, the problem of the
beginning of the Universe in time in the past (problem of
cosmological singularity). The possible solution of DE-problem in
the frame of PGTG was discussed in \cite{20}. Below we will show
that the PGTG offers opportunity to solve also the DM-problem
together with the DE-problem. In Section 2 the cosmological
equations for homogeneous isotropic models (HIM) of PGTG are
given. In Section 3 the restrictions on indefinite parameters of
${\cal L}_{\rm g}$ leading to a possible solution of DM- and
DE-problems in cosmology are obtained. In conclusion some physical
problems in connection with proposed solution are discussed.

\section{Cosmological equations of isotropic cosmology in PGTG}

We will consider the PGTG based on sufficiently general following
expression of gravitational Lagrangian (definitions and notations
of \cite{20} are used below):

\begin{eqnarray}
\label{1}
{\cal L}_{\rm g}=\left[f_0\,
F+F^{\alpha\beta\mu\nu}\left(f_1\:F_{\alpha\beta\mu\nu}
                +f_2\: F_{\alpha\mu\beta\nu}+f_3\:F_{\mu\nu\alpha\beta}\right) \right. \nonumber \\ \left.
        + F^{\mu\nu}\left(f_4\:F_{\mu\nu}+f_5\: F_{\nu\mu}\right)+f_6\:F^2
    \right.  \nonumber
 \\ \left.
    +S^{\alpha\mu\nu}\left(a_1\:S_{\alpha\mu\nu}+a_2\: S_{\nu\mu\alpha}\right)
    +a_3\:S^\alpha{}_{\mu\alpha}S_\beta{}^{\mu\beta}\right].
\end{eqnarray}

The Lagrangian (\ref{1}) includes the parameter $f_0=(16\pi G)^{-1}$ ($G$ is Newton's gravitational constant,
the light velocity $c=1$) and a number of indefinite parameters: $f_i$ ($i=1,2,...6$)
and $a_k$ ($k=1,2,3$). Physical consequences of PGTG depend essentially
on restrictions on indefinite parameters $f_i$ and $a_k$. Some such restrictions will be given below
by investigation of HIM.

In the framework of PGTG any HIM is described by three functions
of time: the scale factor of Robertson-Walker metrics $R$ and two
torsion functions $S_{1}$ and $S_{2}$ \cite{21, 22}. Gravitational
equations for HIM filled with spinless matter with two torsion
functions corresponding to gravitational Lagrangian (\ref{1}) were
analyzed in \cite{20}. These equations allow to obtain
cosmological equations generalizing Friedmann cosmological
equations of GR and equations for torsion functions. Unlike metric
theories of gravity terms of ${\cal L}_{\rm g}$ quadratic in the
curvature tensor do not lead to higher derivatives in cosmological
equations. Higher derivatives can appear because of terms of
${\cal L}_{\rm g}$ quadratic in the torsion tensor. In order to
exclude higher derivatives from cosmological equations we have to
put the following condition for indefinite parameters $a_k$: $2a_1
+ a_2 + 3a_3=0$ \cite{23}. Besides this condition the following
restriction on $f_i$ was used in \cite{20}: $f_2 + 4f_3 + f_4 +
f_5 =0$; by this restriction gravitational equations take a more
symmetric form. Then the cosmological equations for HIM include
three following indefinite parameters: the parameter $\alpha
\equiv f/({3f_0^2 })$ ($f = f_1 + \frac{{f_2 }} {2} + f_3  + f_4 +
f_5  + 3f_{6}$) with inverse dimension of energy density, the
parameter $b = a_2  - a_1$ with the same dimension as $f_0$ and
dimensionless parameter $\veps  = (2f_1 - f_2)/f$. Explicit form
of cosmological equations is the following \cite{20}:

\begin{eqnarray}
\label{2}
    \frac{k}{R^2} + (H-2S_1)^2 =  \nonumber \\
     \frac{1} {{6f_0 Z}}
        \left[
            {\rho  +6\left(f_0 Z- b\right) S_2^2
            + \frac{\alpha }{4} \left( {\rho  - 3p - 12bS_2^2 } \right)^2 }
        \right] \nonumber \\
        - \frac{{3\alpha \varepsilon f_0 }} {Z}
            \left[
                {\left( {HS_2  + \dot S_2 } \right)^2
                + 4\left( {\frac{k}{{R^2 }} - S_2^2 } \right)S_2^2 }
     \right],
\end{eqnarray}

\begin{eqnarray}
\label{3}
    \dot{H}+H^2-2HS_1-2\dot{S}_1 = \nonumber \\
     -\frac{1} {{12f_0 Z}}
        \left[
            \rho  + 3p - \frac{\alpha } {2} \left( {\rho  - 3p - 12bS_2^2 } \right)^2
        \right]
\nonumber\\
        - \frac{\alpha \varepsilon }{Z}\left( {\rho  - 3p - 12bS_2^2 } \right)S_2^2
\nonumber \\
        + \frac{{3\alpha \varepsilon f_0 }} {Z}
            \left[ {\left( {HS_2  + \dot S_2 } \right)^2
                + 4\left( {\frac{k}{{R^2 }} - S_2^2 } \right)S_2^2 }
            \right],
\end{eqnarray}

\noindent where $H=\dot{R}/R $ is the Hubble parameter (a dot
denotes the differentiation with respect to time), $\rho$ is the
energy density, $p$ is the pressure and $Z = 1 + \alpha\left(\rho
- 3p - 12( {b + \varepsilon f_0 } \right)S_2^2)$. According to
gravitational equations the torsion function $S_1$ take the
following form:
\begin{equation} \label{4}
    S_1  = -\frac{\alpha }{4Z} \left[
            \dot \rho - 3\dot p + 36\veps f_0 H S_2^2
            -12\left( {2b - \veps f_0 } \right) S_2 \,\dot S_2
        \right].
\end{equation}
Derivatives of the energy density and pressure can be excluded
from (\ref{4}) by using the conservation law, which in the case of
spinless matter minimally coupled with gravitation takes the same
form as in GR:
\begin{equation}\label{5}
    \dot \rho  + 3H\left( {\rho  + p} \right) = 0.
\end{equation}
The torsion function $S_2$ satisfies the differential equation of the second order:
\begin{eqnarray}
\label{6}
    \varepsilon \left[ {\ddot S_2  + 3H\dot S_2  + (3\dot H  - 4\dot S_1  + 12HS_1
        - 16S_1^2) S_2 } \right]
\nonumber \\
        - \frac{1} {{3f_0 }}\left( {\rho  - 3p - 12bS_2^2 } \right)S_2
        - \frac{{\left( {1 - b/f_0}\right)}} {3 \alpha f_0}S_2  = 0\,.
\end{eqnarray}
Cosmological equations (\ref{2})-(\ref{3}) together with equations (\ref{4}) and (\ref{6}) for torsion functions describe the evolution
of HIM by given equation of state for gravitating matter.

\section{Accelerating Universe without dark energy and dark matter}

Now we will analyze the following question: by what restrictions
on the indefinite parameters cosmological equations of PGTG for
HIM describe the evolution of the Universe in agreement with
actual observations without using notions of dark matter and dark
energy. By taking into account that various parameters of HIM have
to be small at the asymptotics, when values of energy density are
sufficiently small, we see from (\ref{6}), that if $|\veps|\ll 1$,
the torsion function $S_2$ has at asymptotics the following value:
\begin{equation}\label{7}
S_2^2  = \frac{{1  - b/f_0}} {{12\alpha b}} + \frac{{\rho  - 3p}}
{{12b }}.
\end{equation}
Then we have at asymptotes: $Z \to (b/f_0)$, $S_1\to  0$ and the
cosmological equations  (\ref{2})--(\ref{3}) at asymptotics take
the following form:
 \begin{equation}\label{8}
    \frac{k} {{R^2 }} + H^2  = \frac{1} {{6f_0 }}\left[ \rho (f_0/b) + \frac{1}{4} \alpha^{-1}(1 - b/f_0)^2 (f_0/b)
    \right],
\end{equation}
\begin{equation}\label{9}
    \dot H + H^2  =  - \frac{1} {{12f_0 }}\left[ (\rho + 3p)(f_0/b) - \frac{1}{2} \alpha^{-1}(1 - b/f_0)^2 (f_0/b)\right].
\end{equation}
Effective cosmological constant in cosmological equations (8)-(9)
is induced by spacetime torsion function (7). According to
(\ref{8})--(\ref{9}) the evolution of HIM at the asymptotics
depends on two indefinite parameters: $\alpha$ and $b$. Let us to
compare (\ref{8})--(\ref{9}) with Friedmann cosmological equations
of GR:
\begin{equation}\label{10}
    \frac{k} {{R^2 }} + H^2  = \frac{1} {{6f_0 }} \rho_{tot} ,
\end{equation}
\begin{equation}\label{11}
    \dot H + H^2  =  - \frac{1} {{12f_0 }} (\rho_{tot} + 3p_{tot}),
\end{equation}
where $\rho_{tot}$ and $p_{tot}$ are total values of energy
density and pressure including contributions of three components:
baryonic matter, dark matter and dark energy: $\rho_{tot}$ =
$\rho_{BM}$ + $\rho_{DM}$ + $\rho_{DE}$, $p_{tot}$ = $p_{BM}$ +
$p_{DM}$ + $p_{DE}$. In the case of standard $\Lambda CDM$-model
one uses at present epoch for baryonic and dark matter the
equation of state of dust ($p_{BM}=p_{DM}=0$), and for dark energy
$p_{DE}=-\rho_{DE}$. According to observational data, the Universe
evolution is in agreement with Friedmann cosmological equations
(\ref{10})-(\ref{11}) for flat model ($k=0$), if one supposes that
the contribution of dark matter and dark energy to energy density
of the Universe approximately is the following: $\rho_{DM0}=0.23
\rho_{cr}$, $\rho_{DE0}=0.73 \rho_{cr}$, where $\rho_{cr} = 6f_0
H_0^2$ and values of physical parameters at present epoch are
denoted by means of the index "0". By comparing cosmological
equations (\ref{8})-(\ref{9}) with (\ref{10})-(\ref{11}) we see
that cosmological equations of PGTG at asymptotics have
quasi-Friedmannien structure and lead to the same consequences as
Friedmann cosmological equations of GR, if one supposes that the
energy density $\rho$ in (\ref{8})-(\ref{9}) corresponds to all
physical matter in the Universe, which is practically equal to
baryonic matter (by supposing that the contribution of invisible
non-baryonic matter in the form of neutrino etc is sufficiently
small), by certain values of parameters $b$ and $\alpha$, namely
if $b = f_0 (\rho_0/(\rho_0 + \rho_{DM0}))$ and $\alpha =
\frac{1}{4} {\rho_{DE0}}^{(-1)} (1 - b/f_0)^2 (f_0/b)$. Obtained
estimation of $\alpha$ corresponds to energy density of order of
average energy density in the Universe at present epoch and
differs from estimation of $\alpha$ used in our previous papers
\cite {17,18,19,20, 23}, where the value of $\alpha$ corresponds
to the scale of extremely high energy densities at the beginning
of cosmological expansion. Previous estimation of $\alpha$ was
obtained by investigation of HIM with the only torsion function
$S_1$, and it was introduced in order to satisfy the
correspondence principle with GR by description of such HIM at
asymptotics, where values of energy density are sufficiently
small. However, as follows from our consideration given above,
such estimation is not necessary in the case of HIM with two
torsion functions. Moreover, in the case of obtained estimation
for parameters $b$ and $\alpha$ the fine tuning problem of
defining of $b$ in \cite{20} disappears. Note that terms in
cosmological equations containing the parameter $b$ describe the
change of gravitational interaction provoked by spacetime torsion.
When effective cosmological constant in (\ref{8})-(\ref{9}) was
small in comparison with the first term in the right hand side of
(\ref{8})-(\ref{9}), gravitational attraction was larger in
comparison with GR and Newton's theory of gravity. This fact could
play the important role by formation of the large scale structure
of the Universe ensuring additional gravitational attraction,
which is provoked in the frame of standard theory by the dark
matter. However, now when effective cosmological constant
dominates, gravitational interaction has the repulsive character
and leads to acceleration of cosmological expansion. Note that if
some non-baryonic invisible matter exists and gives certain
contribution to energy density $\rho$ in cosmological equations
(\ref{8})--(\ref{9}), in this case obtained estimation for
parameters $b$ and $\alpha$ will be changed.

Cosmological equations (\ref{8})--(\ref{9}) are valid only in the
zeroth approximation with respect to small parameter $\veps$. It
is interesting to analyze observational cosmological data by
taking into account corrections connected with $\veps$ in order to
estimate more precisely the role of spacetime torsion at present
epoch of cosmological evolution. By using obtained estimation for
parameters $\alpha$ and $b$ it is interesting also to study HIM at
the beginning of cosmological expansion with the purpose to build
totally regular Big Bang scenario. However, these problems will be
object of our further investigations.

\section{Conclusion}

We see that the PGTG leads to essential changes of gravitational
interaction not only at extreme conditions (extremely high energy
densities and pressures) in the beginning of cosmological
expansion, but also at present epoch. These changes allow us to
explain the cosmological observational data associated in the
frame of GR with the notions of "dark matter" and "dark energy"
without using these notions. From the point of view of considered
PGTG, the notions of "dark matter" and "dark energy" play the role
similar to that of "ether" in physics before the creation of
special relativity theory by A. Einstein. Unlike GR, in the frame
of PGTG Newton's law of gravitational attraction is not applicable
at cosmological scale. If we remember that the "dark matter"
notion was introduced by applying Newton's law of gravitational
attraction at the galactic scales, the problem of the
investigations of inhomogeneous gravitating systems at galactic
scales in the framework of PGTG becomes very actual. Although the
vacuum Schwarzschild solution for the metrics with vanishing
torsion is an exact solution of PGTG for any values of indefinite
parameters of the gravitational Lagrangian (\ref{1}) that allows
us to explain the usual gravitational phenomena in the Solar
system, for the above mentioned restrictions on indefinite
parameters of ${\cal L}_{\rm g}$ the Birkhoff theorem \cite{24} is
not valid. This means that there are other solutions in this case,
and possibly we have to use in the Solar system the solution,
which deviates from the vacuum Schwarzschild solution. (In
connection with this let us to remind about the problem of Pioneer
anomaly). The search for the criteria that allow us to be able to
choose physically acceptable solutions is warranted and also
important for PGTG.

{Acknowledgements}

The author is very grateful to Prof. Jaan Einasto for discussions
of problems presented in this Letter.

\end{document}